\documentclass[twocolumn]{emulateapj}
\usepackage{natbib}
\usepackage{graphicx}
\usepackage[dvipsnames]{color}
\usepackage{float}
\usepackage{amssymb}
\usepackage{longtable}
\usepackage{enumerate}
\usepackage{gensymb}
\usepackage{soul}
\usepackage{amsmath}

\shorttitle{Dipole Qualifier of EoR Signal, and Foreground Estimation}
\shortauthors{Deshpande}

\begin{document}

\title{Dipole Anisotropy as an Essential Qualifier for the Monopole Component 
of the Cosmic-Dawn Spectral Signature, and the Potential of Diurnal Pattern for Foreground Estimation}

%BLOCKED%\correspondingauthor{Avinash Deshapnde}
%BLOCKED%\email{desh@rri.res.in}

%BLOCKED%\author[0000-0002-5146-2163]{Avinash A. Deshpande}
\author{Avinash A. Deshpande}
\affil{Raman Research Institute
C. V. Raman Avenue, Bangalore 560080, India; desh@rri.res.in\\}

\begin{abstract}

While the importance of detecting 
the Global spectral signatures of the red-shifted 21-cm line of 
atomic hydrogen from the very early epochs cannot be overstated,
the associated challenges primarily include  
isolating the weak signal of interest from the orders of magnitude 
brighter foregrounds, and extend equally to reliably
establishing the origin of the {\it apparent} global signal 
to the very early epochs.
This letter proposes a critical dipole 
test that the measurements of the monopole 
component of the spectrum of interest
should necessarily pass. 
Our criterion is based on
a unique correspondence between the intrinsic 
monopole spectrum and the differential spectrum 
as an imprint of dipole anisotropy 
resulting from motion of observer with respect to the rest frame of our source
(such as that of our Solar system, interpreted from the dipole 
anisotropy in CMBR).
More importantly,  
the spectral manifestation of the dipole anisotropy gets {\it amplified}
by a significant factor, depending on the monopole spectral slopes,
rendering it feasible to measure.
We describe details of such a test, 
and illustrate its application with the help of simulations.
The letter also alludes to
a novel model-independent path
toward isolating the foreground contribution, using the
diurnal pattern readily apparent in drift-scan observations. 
Such dipole qualifier for the monopole spectrum, when combined with
reliable foreground estimation, is expected to
pave way for in situ validation of spectral signatures from early epochs,
important to presently reported and future detections of EoR signal.

\end{abstract}


\section{Introduction} \label{sec:intro}

A number of on-going and planned future efforts at low radio
frequencies aim to detect
precious tokens of the yet unobserved details of the transition from the 
dark ages to the cosmic dawn and beyond to completion of 
reionization, heralded by the first stars 
(Bowman et al. 2018, and references therein). 
The potential detectability of global signal from the red-shifted 21-cm line of 
atomic hydrogen across this cosmic transition was first discussed by
Shaver et al. (1999).
Detection of such signals holds unmatched promise to reveal several key 
details of the physical condition 
and constituents of the universe during these early epochs 
(see Pritchard \& Loeb 2012, and references therein).

Based on their most recent spectral measurement in the 
spectral range 50-100 MHz, Bowman et al. (2018, BR3M18 hereafter) 
have reported ``detection of a flattened absorption profile 
in the sky-averaged 
radio spectrum".
The authors point out the key element 
of surprise,  namely, the depth and flatness of the profile are 
significantly higher than even the deepest predicted 
(Cohen et al. 2017). 
Not surprising is of course the nature of reaction stimulated 
by this news, which includes not only a burst of communication on
the implications of this finding, but also the urgency for
competing radiometers globally probing this spectral window
to verify, and possibly confirm, the reality of the reported absorption
profile.

While appreciating the challenges in detection of 
signatures from HI-line at these early epochs, it is worth noting that 
the corresponding Monopole component or the so-called 
Global signal -- manifested as a faint spectral signature -- is 
considered to be relatively readily detectable, if only the native 
radiometer sensitivity in the spectroscopic measurements were 
alone to dictate reliability of the probe (see early discussion in
Shaver et al. 1999). There is little 
reason to doubt that if the origin of signal reported by BR3M18
indeed corresponds to those early epochs, one would expect 
a prompt confirmation of the spectral signature to be forthcoming 
soon from measurements with other radiometers. 
However, the converse can not be stated with matching confidence.

The reasons and the need for due caution have been well appreciated,
and stem from the high magnitude and uncertainty associated with
contamination or confusion from other potential contributors.
The contaminants range from
a wide variety of astronomical sources, bright and faint,
in the foreground, to a set of systematics and variations 
traceable to man-made signals or measuring instruments/setup
(see Shaver et al. 1999, Pritchard \& Loeb 2012; BR3M18; 
and references in the latter).

Despite all the careful accounting and removal of
the obvious and subtle contributions, it still remains 
a significant challenge to 
distill out the underlying EoR Global signal,  
in the presence of
bright foreground emission, even though procedures to 
fit together a suitable spectral model for foreground and 
monopole signature have been routinely employed.
A reliable way to estimate and separate the foreground contribution
is certainly desired.

Ideally, we also require a critical test to reliably verify that
the apparent EoR signal (e.g. as in BR3M18 or any future report)
is indeed from the early epochs.
However, despite an earlier discussion by Slosar (2017)
on the magnitude of the dipole spectral signature, and mention of
its potential use to ``cross-check measurements derived from the
monopole", the dipole spectrum measurements 
have not yet received its due attention.\footnote{
Ironically, the author was unfortunately unaware of this paper till 
after the submission of initial manuscript, which is now revised accordingly,
thanks to Ravi Subrahmanyan drawing attention to this paper.}
Interestingly, Slosar (2017) has argued in favor of measuring the dipole
spectral signature, even if weak, instead of monopole, since the former
would be much less contaminated by galactic foregrounds, but also remarks
``one could imagine an experiment that would measure both at the same time".
 
In this letter, we propose a dipole-based in situ qualifier
that the measured EoR spectra should necessarily pass to be consistent
with being a monopole component of the signal from the early epochs, and
show how this signature can be measured, despite its weakness.
This qualifier has the potential to serve as a conclusive test as well,
and also to provide a useful reciprocal prediction.
We also draw attention to a potentially effective path 
to estimate and isolate the foreground contamination 
in a model-independent manner, based
on apparent diurnal pattern.

\section{Direction dependence of apparent spectrum of EoR signal, 
and amplified diurnal dipole imprint}

\begin{figure}[ht!]
\includegraphics[scale=0.32, angle=-90]{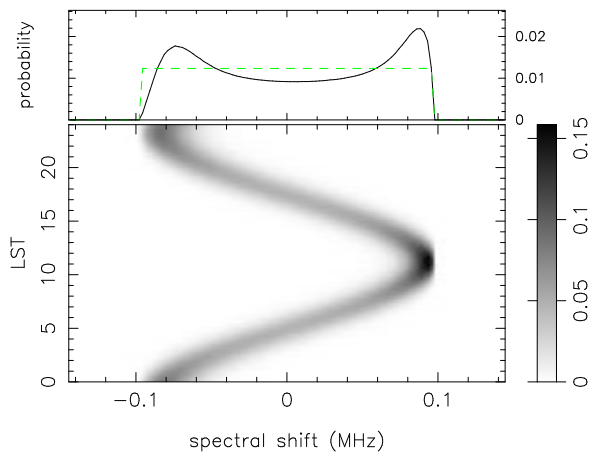}
\caption{Spectral shifts and spread are shown in gray-scale 
(bottom panel) as a function of 
Local Sidereal Time (LST).
The top panel shows the integrated effect of smearing over 
the sidereal day, for the transit observation (solid line), 
as well as for all-sky integration (dash line).
}
\end{figure}
  
\begin{figure}
  \centering
  \begin{tabular}[b]{@{}p{0.45\textwidth}@{}}
    \centering\includegraphics[scale=0.31, angle=-90]{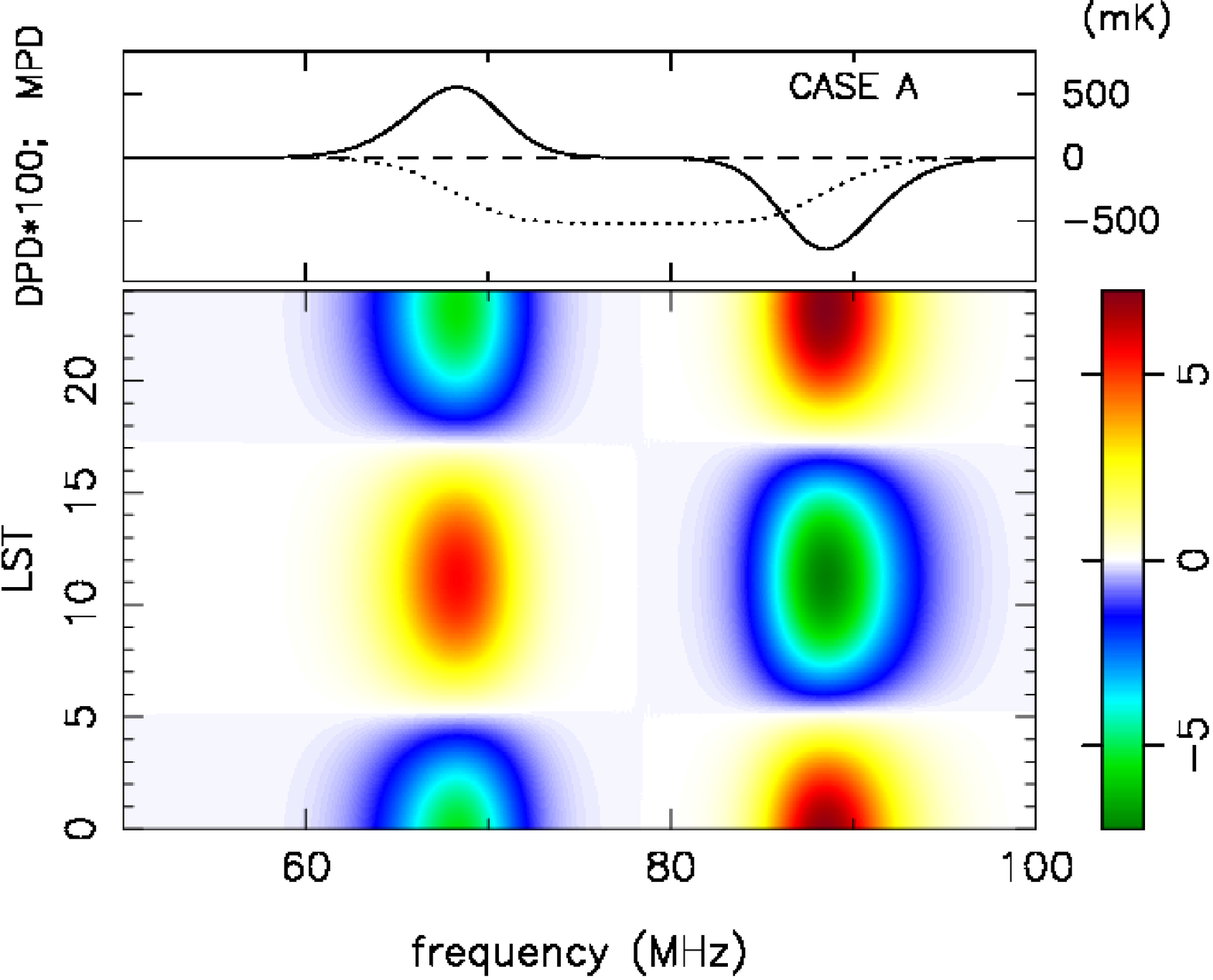} \\
    \centering\includegraphics[scale=0.31, angle=-90]{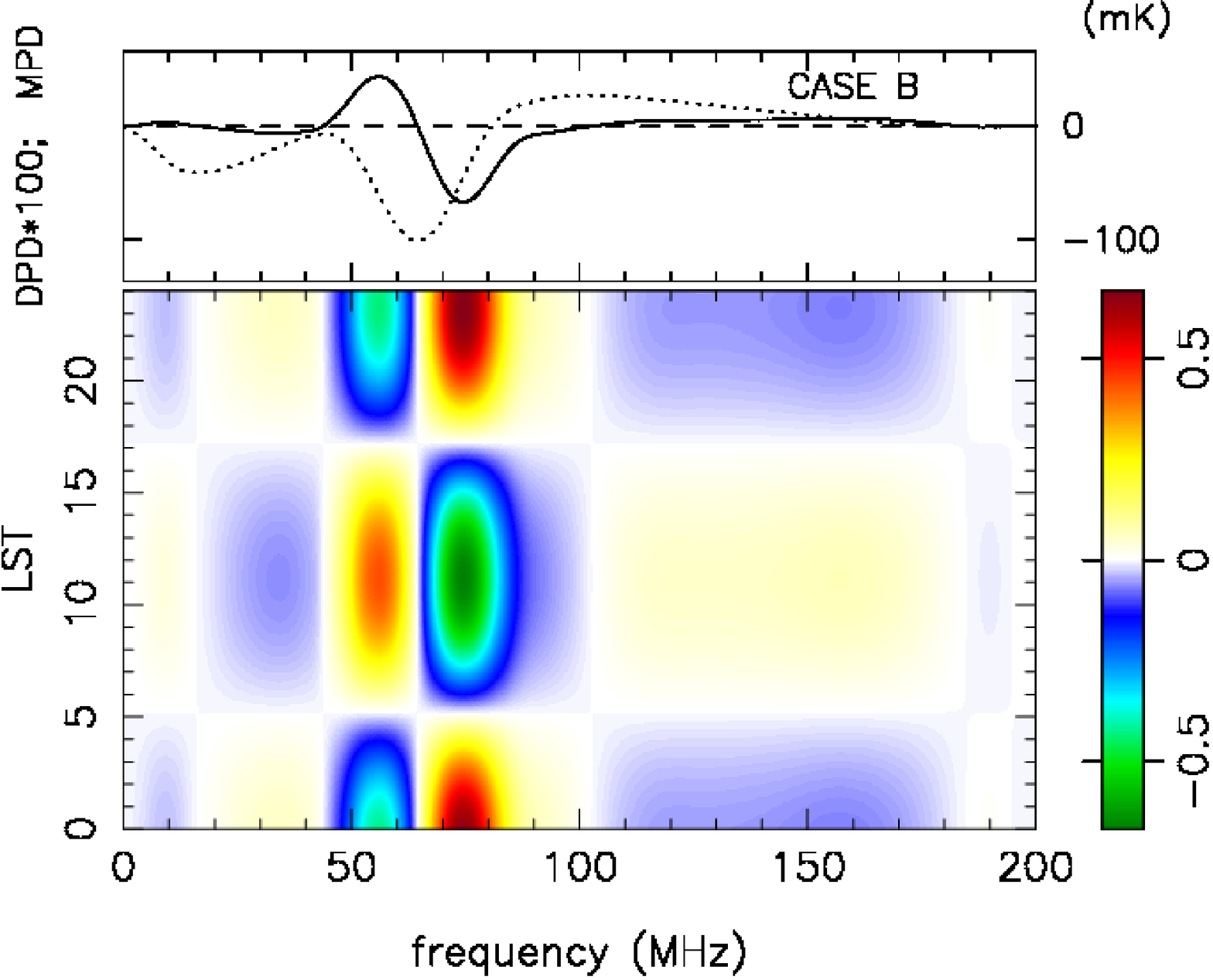} \\
\end{tabular}
\caption{Expected set of residual or difference spectral profiles across the 
entire LST range, after subtraction
of ``all-sky" average spectral profile from the simulated set 
containing the discussed mild dipole frequency modulation, are shown. 
The 24-hour cycle and the correspondence with
the first (spectral) derivative of the assumed underlying monopole spectrum
(of course multiplied by $\nu$, as in Equation 6), are clearly evident. 
The amplitude of the
co-sinusoidal variation (solid line), after scaling up by 100,
is shown in the top panel for each case A \& B, along with the assumed monopole spectrum (dots).}
\end{figure}
  
The generally accepted interpretation for the CMB Dipole Anisotropy
(DA hereafter) is the Doppler effect due to
the peculiar velocity $v$ of the Solar system (= 369.0$\pm$0.9 $km\,s^{-1}$
toward Galactic coordinates
(l,b) = (264$\degree.99\pm$0$\degree$.14, 48$\degree$.26$\pm$0$\degree$.03);
Hinshaw et al. 2009)
with respect to the rest frame of CMB 
(see also, Burigana et al. 2018, and references therein), 
although alternative possibilities
have also been discussed (see for example, Inoue \& Silk 2006).
We explore below how the associated
Doppler effect would manifest in apparent spectral profiles of 
the much sought-after monopole component of EoR signal.

In general, given the approaching velocity $v$, radiation at frequency $\nu$,
reaching from an angle $\psi$ with respect to the direction 
of the velocity, 
will be shifted to apparent frequency $\nu_a$, given by 
$\nu_a = \nu {(1+\beta cos\psi)}/{\sqrt{1-\beta^2}}$
where $\beta = v/c$, and $c$ is the speed of light.
When $v/c \ll 1$, the Doppler shift  
$\Delta \nu =  (\nu_a - \nu)$
will also be proportionally small compared to $\nu$, and
can be approximated to the first order as $\nu \beta cos\psi$.
Thus, for an intrinsic EoR monopole spectral profile $\Delta T(\nu)$,
the apparent deviation profile (usually in units of temperature),
obtained after careful subtraction of foreground and 
appropriate calibration (including CMBR dipole variation,
to avoid its implied amplitude scaling of monopole spectrum, even though small), 
would be direction dependent (in scaling of its spectral axis),
and can be expressed as 
$\Delta T_a(\nu,\psi) = \Delta T(\nu(1 - \beta cos\psi))$,
or more generally, 
\begin{equation}
\Delta T_a(\nu,\hat{s}) 
= \Delta T(\nu(1 - \beta \hat{s}.\hat{s}_{DA})) 
\end{equation}
where $\hat{s}$ and $\hat{s}_{DA}$ (or RA$_{DA},\delta_{DA}$)
are unit vectors in the directions of
source and DA, respectively, and ``." indicates dot product of 
these vectors. A sky-averaged version of the apparent spectrum, 
when integrated over full sky (4$\pi$ sr), 
the contribution at each $\nu_a$ 
can be shown to be an average of the underlying spectrum $\Delta T$ over 
a window $\nu\pm\nu\beta$, amounting to smoothing by a rectangular spectral 
window of width proportional to $\nu$. Thus,
\begin{equation}
<\Delta T_a(\nu)>_{all-sky} = \frac{1}{2\nu\beta}\int_{-\nu\beta}^{+\nu\beta} \Delta T(\nu + f) df
\end{equation}

\begin{figure}
  \centering
  \begin{tabular}[b]{@{}p{0.45\textwidth}@{}}
    \centering\includegraphics[scale=0.40, angle=-90]{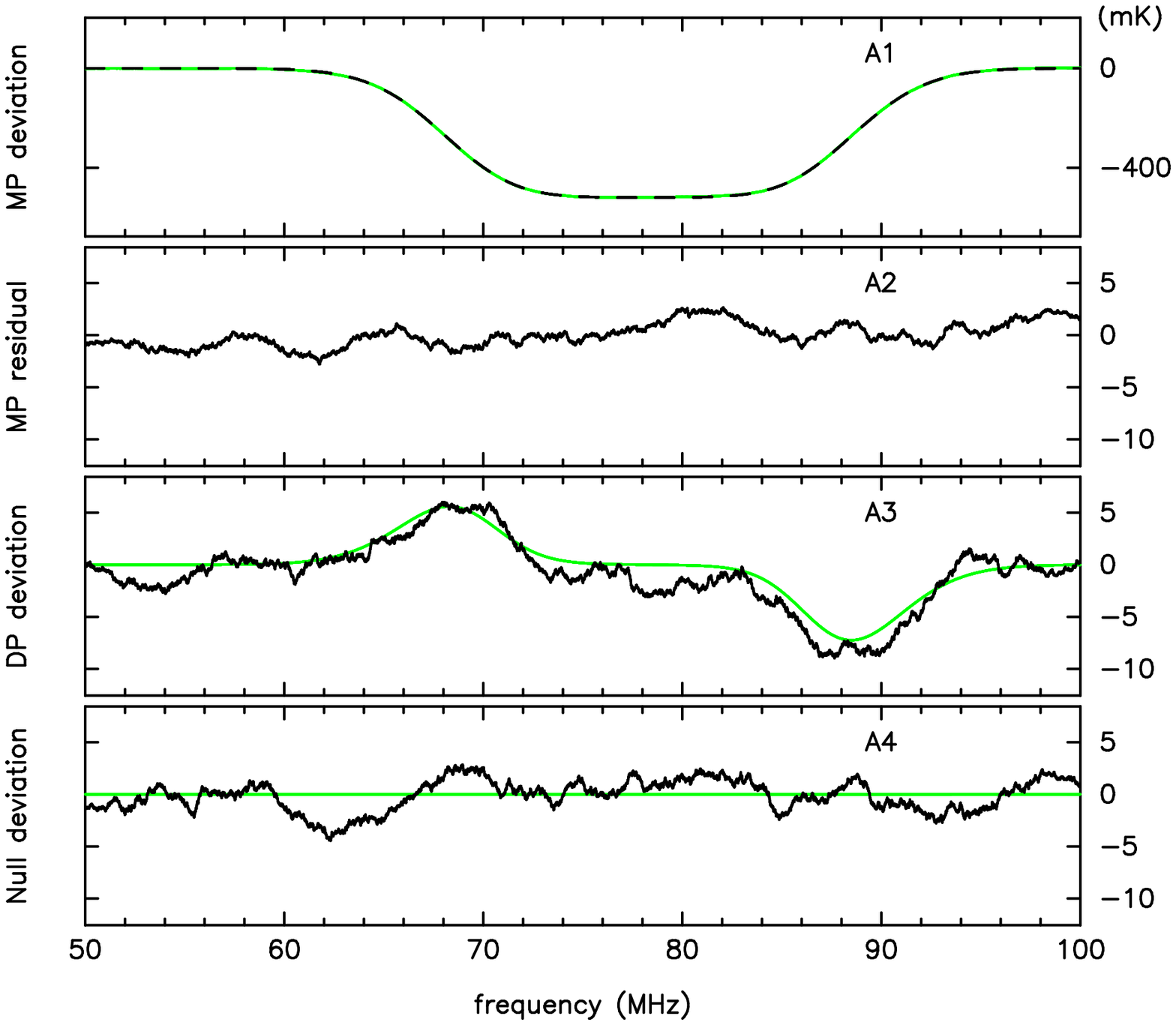} \\
    \centering\includegraphics[scale=0.40, angle=-90]{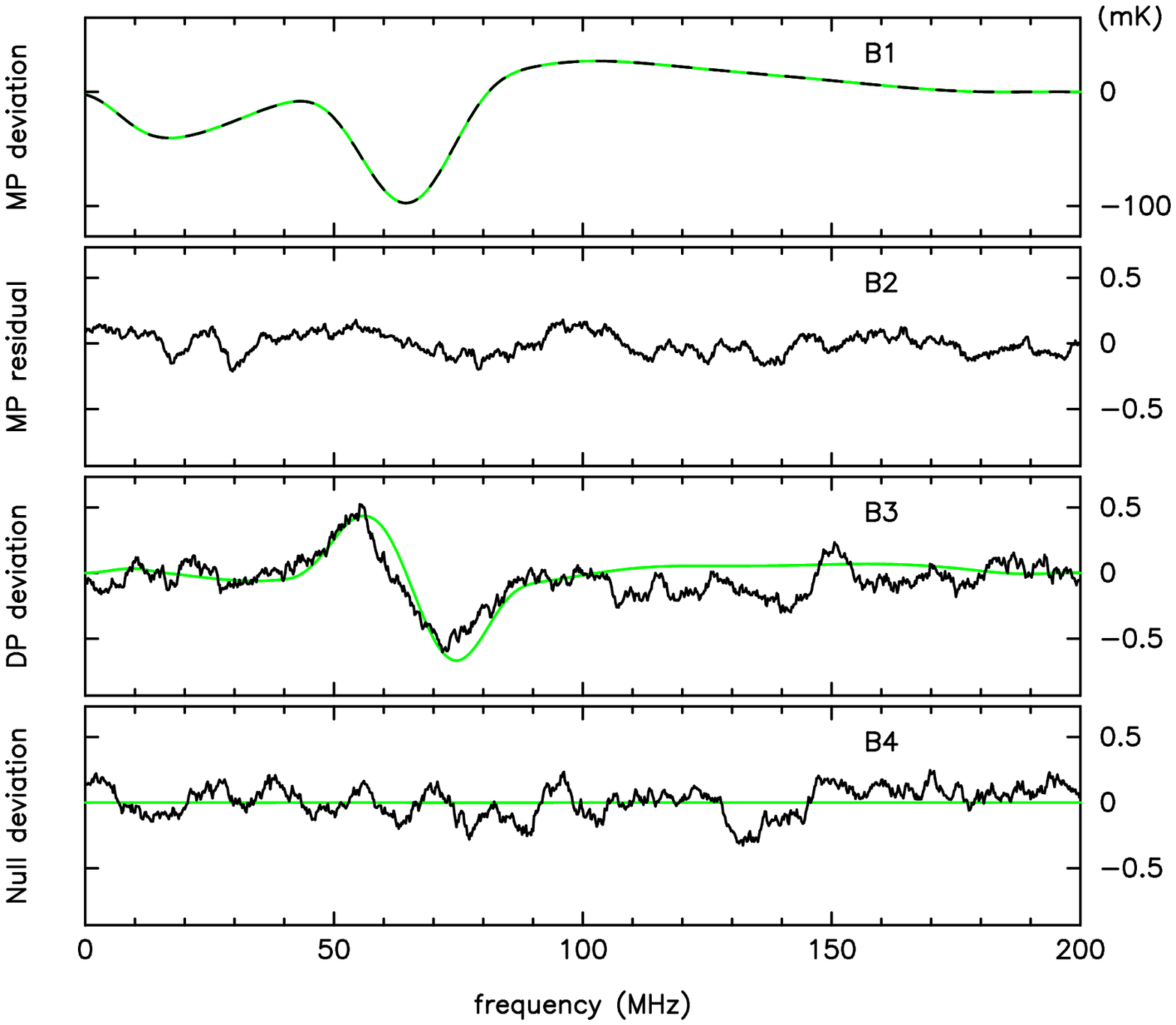} \\
\end{tabular}

\caption{Resultant spectra from analysis of a simulated 
dynamic spectrum across the full range of LST, for each
of the two monopole profiles considered, case A \& B,  
are shown separately in the upper and lower halves, respectively.
In each half, 
the top panel (A1, B1) shows an assumed intrinsic spectral 
profile (green) along with the average of the simulated set of spectra
(dash line), representing an `observed' average profile.
These two profiles are indistinguishable on the
compressed scale, and hence the difference is separately shown
in the panel below (A2, B2). The second 
panel from bottom (A3, B3) shows the 
estimated spectrum of the amplitude of the co-sinusoidal variation across 
the sidereal cycle referenced to the LST of the dipole (black), along 
with the expected spectral profile (color). Similarly, the bottom 
panel (A4, B4) shows a dipole profile extracted 
with orthogonal modulation,
along with an expected null profile, for reference.}
\end{figure}
  
For $\beta \ll 1$, the smoothing is not expected to be noticeable,
except for unlikely sharp features in the underlying spectrum,
particular since there is no net shift or stretching/contraction
of the profile, given symmetry
of the window about zero shift. 

In practice, even in a snap-shot measurement, 
sky signals (i.e. the associated power spectra) 
from a range of directions 
would be averaged over the visible sky, weighted
by the instrumental angular response $G(\hat{s})$, or $G(\theta,\phi)$ 
as a function of azimuth $\phi$ and zenith angle $\theta$, 
with a nominal pointing center, say,  $\hat{p}$, expressible in terms of
Right Ascension (or Local Sidereal Time) and declination, (RA,$\delta$) or (LST,$\delta$). 
For radiation at a given radiation frequency $\nu$,
the beam-averaging would result in a spread or smear in the apparent  $\nu_a$, 
given by,
\begin{eqnarray}\nonumber
\Delta T_{obs}(\nu_a,\hat{p}) &=&\\ 
\frac{\int_{\theta} \int_{\phi} G(\theta,\phi) \Delta T(\nu, \hat{s}) sin\theta d\theta d\phi}{\int_{\theta} \int_{\phi} G(\theta,\phi) sin\theta d\theta d\phi}
\end{eqnarray}
such that $\nu_a = \nu(1 + \beta \hat{s}.\hat{s}_{DA})$,
and the direction $\hat{s}$ is a function of $\theta,\phi$.
Figure 1 shows an example of how such spread in the shift $(\nu_a - \nu)$ 
would vary as a function of LST, 
computed for $\nu$ = 78.3 MHz, and
assuming meridian transit observations with a 30$\degree$ (FWHM)
beam pointed to the zenith at a latitude of -26$\degree$.7 (chosen to be 
similar to the observing setup of BR3M18, except that we assume the beam to
be frequency-independent, for simplicity).
The profile of the spectral spread, or equivalently a smoothing function, 
also show significant variation as a function of LST, as expected.
 
Given an  observing band from $\nu_{min}$ to $\nu_{max}$, 
if even the maximum possible shift $\nu_{max} \beta$ is  $\ll$ 
width of the narrowest feature in the profile, it is easy to 
see that the residual profile deviation $\delta T_a(\nu,\hat{s})$ 
from the original signal profile $\Delta T(\nu)$, 
or say, the difference profile,
can be expressed as
\begin{eqnarray} %\nonumber
\delta T_a(\nu,\hat{s})  
&=& \Delta T(\nu(1 - \beta \hat{s}.\hat{s}_{DA})) - \Delta T(\nu)\\ 
&=& -\nu \beta (\hat{s}.\hat{s}_{DA})(d(\Delta T(\nu))/d\nu)\\ 
&=& -\nu \beta (\hat{s}.\hat{s}_{DA})D(\nu) 
\end{eqnarray}
where $D(\nu) = d(\Delta T(\nu))/d\nu$, 
the first derivative of the original profile with respect to frequency $\nu$,
and 
\begin{equation}
\hat{s}.\hat{s}_{DA} = \cos\delta \cos\delta_{DA}\cos(LST-RA_{DA})
+ \sin\delta\sin\delta_{DA}
\end{equation}

The essential origin, and the implied magnitude, of this variation 
induced by observer motion are no different from that discussed by 
Ellis \& Baldwin (1984), although described in the context of
apparent source distribution and related parameters.
The major difference is that 
here the so-called ``spectral index" is neither small nor ``constant",
thanks to the expected variations across the spectrum associated
with the red-shifted HI from early epochs, and the associated 
spectral slope $D(\nu)$.
The relatively rapid and large magnitude changes
in $D(\nu)$ not only make the difference profile $\delta T_a(\nu,\hat{s})$
spectrally featureful, but also correspondingly amplify
its variation as function of direction $\hat{s}$ (or LST).
In vivid contrast to 
most of the manifestations, of the
dipole anisotropy resulting from the Solar system motion, being typically
of the order of $v/c$, that is a part in thousand, 
the mentioned amplification is found to raise the scale of profile changes 
by typically an order of magnitude, to a percent level.

We illustrate in Figure 2 how the dipole modulation would reveal itself
in the difference profiles across a sidereal day, in an observing setup
similar to that assumed in Figure 1. The cases A and B correspond respectively 
to BR3M18 best-fit profile (50-100 MHz) and the much discussed
theoretically predicted spectrum up to 200 MHz (the ``turning points'' 
data taken from Pritchard \& Loeb 2012).
The simulated set consists of a dynamic spectrum spanning
a sidereal day with 400 time bins, each representing 
an apparent spectrum associated with an assumed intrinsic spectral 
profile as shown in the top panel (dash line), for a time duration of
3.6 minutes per snap-shot, and integrated over the sky 
area defined by the angular response of the instrument.

\section{Dipole qualifier for in situ validation of the monopole component of the EoR signal}

Encouraged by the amplified manifestation of the induced dipole anisotropy
in the apparent spectra of the expected monopole signal from the very early epochs,
we now proceed to propose a critical in situ test to verify its desired origin.

Here, we assume usually recommended sky drift observations 
made with a fixed beam, for simplicity and preferred coherence 
in the data set. Such data are assumed to be in form of an average 
dynamic spectrum, well-calibrated
for system response to the extent possible,
and over a span of one sidereal day
(averaged synchronously over this period, if from multiple days).
 
To ensure that the dipole component of the diurnal pattern
is devoid of any contamination 
from monopole-like contribution
(not necessarily limited to EoR signal)\footnote{Slosar (2017) expression 
for dipole signature contains this avoidable leakage from
the EoR monopole profile.}, we need to calibrate (divide) 
the dynamic spectrum by a factor $ (1 + \beta \hat{s}.\hat{s}_{DA})$, where 
$\hat{s}$ corresponding to the $RA,\delta$ of sky transiting at the zenith.

The next step involves well-recognized challenges,
wherein the foregrounds 
are estimated as well as possible and removed. 
The further processing steps would be obvious to an expert,
but are mentioned below merely for completeness.

a) Averaging the spectra across the entire LST range to obtain 
the monopole spectrum (though most smeared), as an estimate of $\Delta T(\nu)$, 
and subtracting it from the entire set of
spectra to obtain a set of difference profiles  
$\delta T_a(\nu_a,$LST).
b) Extracting the amplitude profile, $\delta T_{dp}^{O}(\nu)$, 
at the fundamental frequency 
of the diurnal variation, using 1-d Fourier transforms along LST axis, 
and rotating the phase to reference it to RA$_{DA}$, 
or by simply filtering the variation with $\cos\Theta$, to obtain
a spectrum potentially containing the dipole signature, as well as with
$\sin\Theta$, where $\Theta = 2\pi($LST-RA$_{DA})/24$ and 
LST,RA are in units of hr. 
The latter result, say $\delta T_{null}^{O}(\nu)$, 
serves as a useful reference profile for
assessing uncertainties in the former\footnote{In what may be a mere coincidence,
nonetheless intriguing, the dipole direction is close to the Galactic pole, 
and hence the transits of Galactic plane occur preferentially close to 
the null of the dipole imprint, leading to reduced contamination from any 
residual foreground due to the plane. In contrast, the reference profile 
gets almost unattenuated contribution from the plane, and provides 
a measure of potential contamination in the dipole spectrum.}. 
When estimated from the original diurnal pattern (without 
removal of foreground, but after due calibration), this reference profile
might serve as a useful initial model of foreground contamination in the
{\it dipole profile}, after appropriate 
scaling (assessed in the spectral region devoid of 
expected dipole signature). 
c) Using the best-fit profile for the monopole component spectrum 
$\delta T_{mp}^{M}(\nu)$, 
and computing a differential profile (first derivative) times frequency $\nu$
as a model profile $\delta T_{dp}^{M}(\nu)$ for the induced dipole component.
d) Cross-correlating or matched-filtering 
$\delta T_{dp}^{O}(\nu)$ with $\delta T_{dp}^{M}(\nu)$ 
to assess significance of the match.
 
Figure 3 illustrates application of above mentioned procedure 
to a simulated set of profiles spanning the entire LST range, 
and for two descriptions of monopole components, 
similar to those in Figure 2, but now with Gaussian random noise
added. The noise level is chosen such that the r.m.s. noise in the
average monopole profile would be 3 and 0.4 mK in the case A and B, 
respectively. The extracted monopole and dipole component profiles
show desired correspondence with the respective expected spectra 
(shown in green), within the noise deviations, which are also found 
consistent with the integration over the entire LST range
and across frequency (smoothing function width of 4 MHz and 8 MHz for 
case A and B, respectively).
 
A few key advantages of the suggested method are worth emphasizing.
Any time-independent contamination 
in the apparent monopole spectrum 
does not contribute to the dipole signature.
Hence, the extracted dipole profile can be expected to be free of
any error in model of the monopole profile, and also
any effectively additive ``local" contributions, 
such as ground pick up, instrumental noise, and
also small multiplicative effects, e.g. remaining systematics from 
inadequate calibration of instrumental response, residual 
spectral modulation due standing waves from reflections, etc.

We have assumed that the spectral profile set we start with
is after foreground removal and calibration. However, since 
(different) foregrounds also will be Doppler modulated differently 
across LST, they can potentially contaminate the dipole spectrum 
of interest, when foreground removal is imperfect.
However, residual contribution, if any, from these across 
the difference spectra would still be smoothly varying,
and might even be proportional to $\nu (1+ \alpha)$ 
(see Ellis \& Baldwin 1984).
Given its smoothness, combined with {\it its presence even 
in spectral regions devoid of monopole/dipole signal}, 
removal of the {\it baseline} may be possible
by modeling with low-order polynomials,
or using models with a few parameters,
even allowing for slow changes in the spectral index $\alpha$
with frequency.
If the above assumptions are rendered 
invalid, or there would be risk of
{\it absorbing} the dipole signature in the fits, a following 
procedure to extract the dipole signature may be employed, once the
model profile of monopole is known and is to be qualified.
A 1/0 mask, say $m(\nu)$ corresponding to the zeroes to be expected in the
implied dipole profile (based on the model monopole profile), and similarly
$M(\tau)$ corresponding to the zeroes in the Fourier transform (FT) 
of the predicted dipole profile are noted\footnote{It is easy to appreciate
that the exact relation between the dipole signature and 
the monopole spectrum continues to hold even for their Fourier transforms.}
An iterative application of these masks in respective domains
on successive forward and inverse Fourier transforms is expected to
converge, resulting in a profile consistent with the provided constraints.
Such filtering of the associated dipole component
benefits from its nulls\footnote{Samples in the input 
profile, and its FT, at the locations of these respective nulls 
(one such amounting to integral of the spectral profile) correspond to 
foreground alone, and may be used as constraints
while modeling the foreground instead.},
which outnumber the order of the
polynomial or number of parameters were to be fitted
in traditional approach. Note that the locations of these
nulls do not change with LST. With profiles at each LST filtered
in this manner, one now looks for the diurnal pattern as a test
of the expected dipole signature in both frequency and LST together.

The true monopole spectrum is not known a priori, and
an apparent monopole spectrum is estimated 
by averaging the corresponding data across the observing span.
It is easy to see that such monopole spectrum will contain also
the contribution associated with the LST-independent term 
(second term in Equation 7), defining a tiny leakage of
the dipole component in to the apparent monopole spectrum.
For $\delta_{DA}\approx-7\degree$, this leakage is rather small,
more so when $|\delta| \ll 90\degree$, a situation
preferred in any case for maximizing the dipole modulation
as far as possible.

The most exciting prospect is of predicting the monopole
spectral profile based on extracted dipole spectrum,
by scaling the model profile, best fit to the latter,
by $1/\nu$, followed by integration.
Comparison of this derived version with the observed monopole
profile would provide unprecedented scrutiny of the fidelity of
the latter, 
and is likely to be more instructive (relative to the comparison 
suggested in step d)), given the relative immunity
of the dipole profile to contaminants.   
How far this exciting prospect can be realized in reality
remains to be seen, in light of the known challenges in 
reliable estimation of the contaminants.

\section{On the prospects of using diurnal variation for 
in situ estimation of foreground}

Here, we enquire what can be learned about foregrounds themselves
from their significant fraction manifesting 
diurnal variation apparent in the average dynamic spectrum
considered above.
This fraction need not be constant across frequency, particularly
if the angular response depends on frequency, and for other intrinsic reasons.
As already noted, in the Fourier transform (FT) of the data
along time, the sum of monopole signature and average foreground
would define the profile the zero fluctuation frequency, and 
the immediate FT component\footnote{
It is worth pointing out that these Fourier components 
are exact equivalent of the visibilities at the various
spatial frequencies (namely, u,0) for sky
modulated by the system response in declination, 
which one would have wanted to measure for the present and other reasons, 
and are not trivial to measure otherwise from the Earth, without needing to
solve for them from interferometric measurements.} 
at $1/day$ would contain the dipole component of EoR
as well as that of the sky scanned. It is 
the FT component at $2/day$ that is free of both the monopole and dipole, and
which could have been
appealed to provide an estimate of the varying component of
the foreground, if the latter were to be
only a single peak with a width of fraction of a day, say, $\Delta day < 1/3$,
implying coherence across $1/\Delta day$ in fluctuation frequency.

In reality,
the level of coherence across 
different FT components (close to zero fluctuation frequency)
can be significantly low, limiting their utility for the above purpose.
Nonetheless, we can ask
if at least the interrelations between the relevant statistical attributes 
of the varying component 
would be similar or at least smoothly varying across frequency.
The latter is more likely to be true, given the inherent level of
smoothness, commonality of origin,
and the one-sidedness of the foreground intensity distribution.
Note that possible imprint of the instrumental spectral response
will be common to all relevant apparent attributes, and hence
would not be expected to affect their interrelation.

We have assessed this expectation through simulations\footnote{More 
detailed account of the exploration is
beyond the scope of this letter, and will be reported elsewhere.}
 and found that 
the statistical property, such as the 
mean intensity and the standard deviation (from the mean) 
show reliable degree of correspondence, with their ratio showing
only smooth variation, if any, across frequency, except in monopole/dipole
region. Much more work would be needed to explore this aspect further,
and at this stage, we merely wish to draw due attention to
this potentially important possibility 
of in situ estimation of foreground contamination that the statistics 
and components of observed diurnal pattern might offer.

\section{Discussion and Conclusions}

In illustrating application of our method
to the model profile of BR3M18, we have deliberately
assumed a much reduced 
r.m.s. noise (3 mK) in our simulation (case A) compared to 
their reported noise r.m.s. ($\sim$20 mK), to aid ready
detection visually, resulting in 1 mK r.m.s. noise after
4 MHz smoothing. 
While using cross-correlation or matched-filtering to assess
presence of the dipole component, it worth noting that
the sensitivity benefits significantly 
from the effective bandwidth of the dipole
pattern, not limited by the fine resolution of the spectrum.
In the present case, this effective bandwidth would 
be about 8 MHz.
Needless to stress that
for validation of their reported detection using the dipole test, 
the EDGES (BR3M18) spectra would need improved signal-to-noise ratio,
at least by a factor of about $\sqrt5$,
for a 3-$\sigma$ detection of the 
dipole component of 7 mK, implied by their model monopole profile.

The needed sensitivities do appear feasible\footnote{Although the
residual r.m.s. (shown in the Extended Data Figure 9
of BR3M18) is seen to depart significantly from the reference expectation 
(even approaching saturation), the departure appears to be most likely 
a reflection of required
refinement in the presently fitted models (as was the case
before inclusion of the 21-cm model),
and fortunately, there appears to be no indication yet
of any {\it red} process dictating the residuals.
While significant refinement in the models,
facilitated by necessary reduction in the random noise,
may be looked forward to, any errors in the model of the 21-cm
monopole and that of the {\it sky-averaged} foreground, would not affect
the sensitivity of detection for the dipole component, except a
corresponding revision, if any, in its profile prediction.
Subject to the validity of the above understanding,
it would not be surprising if more integration
by a factor of 5 or so in the measurements of BR3M18
were to suffice to provide desired improvement in the sensitivity
to facilitate the necessary refinement in the modeling of systematics,
essential before
assessing possible presence of the dipole signature.}, 
in view 
of {\it also} some of the encouraging on-going efforts (see the list,
in BR3M18, of radiometers that can help verify their finding).

Mutual consistency between the observed monopole spectrum
and the extracted dipole spectrum thus suggests an essential 
and unique in situ test we desire the measurements to pass, 
before the detected signal can be
justifiably viewed as from early epochs. When the consistency
is high enough, the suggested test has the potential to
be even a sufficient criterion.

We wish to also point out in passing 
that the discussed spectral imprint of the
dipole anisotropy has interesting reciprocal implications
for the signature  to be expected across the longitudinal
component of the spatial frequency $k_{||}$, relevant to
probe of the statistical signature of EoR through measurements
of spatial power spectrum at low radio frequencies (for example,
see Datta et al. 2010 for details on such probes), 
and would be rewarding to explore.

In the discussion/illustrations so far, 
we have used the Solar system velocity 
as implied by the CMBR dipole anisotropy, as a conservative estimate. 
It is not known yet if DP anisotropy evolves with red-shift, although
there have been intriguing indications (see for example, Singal 2011). 
In any case,
ready application of our method to dynamic spectrum
in $\nu$-LST plane, 
combined with models for dipole evolution 
as a function of redshift (i.e. presumably smooth spectral dependence of 
$\beta$, RA$_{DA},\delta_{DA}$, if any)
promises worthy tomographic exploration of the dipole imprint,
but only after the primary challenges posed by 
contaminants are met successfully.

\acknowledgments
We thank our referee for constructive criticism and valuable comments,
which helped in improving the manuscript.

\end{document}